\documentclass{emulateapj}

\newcommand{\num}{430}
\newcommand{\tb}{$0.12\,\mbox{K}$}
\newcommand{\jybeam}{$0.15\,\mbox{Jy/Beam}$}
\newcommand{\vlsr}{\ensuremath{v_{LSR}}}
\newcommand{\kms}{\,\mbox{km\,s$^{-1}$}}
\newcommand{\velsel}{$|\vlsr| \ge 90\kms$}
\newcommand{\hb}{\ensuremath{H\beta}}
\newcommand{\hc}{\ensuremath{H\gamma}}
\newcommand{\hd}{\ensuremath{H\delta}}
\newcommand{\ca}{\ion{Ca}{2}}
\newcommand{\ovi}{\ion{O}{6}}
\newcommand{\HI}{\ion{H}{1}}
\newcommand{\fti}{5} 
\newcommand{\lti}{148} 
\def\kpc{\ifmmode{\,\mbox{kpc}}\else\,kpc\fi}

\shorttitle{FHB Stars aligned with HVCs}
\shortauthors{Thom, Gibson \& Christlieb}

\begin{document}

\title{A Catalogue of Field Horizontal Branch Stars Aligned with High
  Velocity Clouds}

\author{Christopher Thom\altaffilmark{1} and Brad K. Gibson\altaffilmark{1}}
\affil{Centre for Astrophysics and Supercomputing, Swinburne University of
  Technology, PO Box 218, Hawthorn, Victoria, 3122, Australia}

\author{Norbert Christlieb\altaffilmark{2}}
\affil{Hamburger Sternwarte, Universit\"at Hamburg, Gojenbergsweg 112,
  D-21029 Hamburg,Germany}
\email{cthom@astro.swin.edu.au}

\begin{abstract}

  We present a catalogue of \num~Field Horizontal Branch (FHB) stars,
  selected from the Hamburg/ESO Survey (HES), which fortuitously align with
  high column density neutral hydrogen (\HI) High-Velocity Cloud (HVC) gas.
  These stars are ideal candidates for absorption-line studies of HVCs,
  attempts at which have been made for almost 40 years with little
  success. A parent sample of 8321 HES FHB stars was used to extract \HI\
  spectra along each line-of-sight, using the \HI\ Parkes All-Sky Survey.
  All lines-of-sight aligned with high velocity \HI\ emission with peak
  brightness temperatures greater than $120\,\mbox{mK}$ were examined.  The
  \HI\ spectra of these \num~probes were visually screened and
  cross-referenced with several HVC catalogues.  In a forthcoming paper, we
  report on the results of high-resolution spectroscopic observations of a
  sample of stars drawn from this catalogue.

\end{abstract}


\section{Introduction}

Discovered by \citet{muller-etal-63-HVC-discovery} over 40 years ago,
High-Velocity Clouds (HVCs) are \HI\ clouds with velocities that do not
conform to simple models of Galactic rotation \citep[e.g.][and references
therein]{wakker-vanwoerden-97}. Since their discovery, HVC distances (and
origins) have been the source of much debate, with various scenarios
suggested.  \citet{shapiro-field-76-ISM} proposed the existence of a
``Galactic Fountain'' to explain the observed soft X-ray background and
\ovi\ absorption lines. This process, in which hot gas is driven up into
the halo by supernovae and stellar winds, which then condenses and falls
back onto the disk, was linked by \citet{bregman-80} to the HVCs \citep[see
also][]{deAvillez-00-galfountain}.  Others have suggested they may be the
result of tidal disruption of satellite galaxies, e.g. the Magellanic
Stream \citep{putman-etal-03-mshvc}. Perhaps they are the infalling source
of star formation fuel long required by chemical evolution models
\citep[e.g.][]{fenner-gibson-03}?  \citet{oort-66-HVC-origins} first
suggested that some subset of HVCs may be at much larger distances, being
the remnants of the galaxy formation process \citep[a suggestion revived by
][]{blitz99}.  Other authors have proposed that the smaller, isolated
Compact HVCs (CHVCs) may be explained under this scenario
\citep[e.g.][]{deheij-etal-02-allsky}, suggesting that such objects may
also be observable around M31 with sensitive \HI\ observations. Subsequent
observations toward M31 have successfully detected HVC-analogs
\citep{thilker-etal-04-M31}, whilst others have failed to find them in
searches of Local Group analogues \citep{pisano-etal-04}.

It is clear that to establish a definitive theory describing HVCs the
largely unknown distances are a key element.  Many physical properties
scale with the distance; physical size ($\varpropto d$), mass ($\varpropto
d^2$), density ($\varpropto d^{-1}$) and pressure ($\varpropto d^{-1}$) for
example. Since HVCs do not conform to models of Galactic rotation, we
cannot use those models to derive distances from their observed velocities.
Hence another method is required.  A distance limit may be set by
high-resolution absorption spectroscopy
\citep{danly-etal-93-ComplexM,schwarz-etal-95,vanwoerden-etal-99-nature}.
If a star is located behind the HVC, the stellar spectrum should contain
the imprint of the HVC at the (known) velocity of the cloud.  A
sufficiently high resolution spectrum should be able to resolve such
features.  Detection of these absorption lines indicates the star is behind
the cloud, and an upper distance limit is set, namely the distance to the
star.

The interpretation of non-detections is more challenging and is succinctly
summarised by \citet{wakker-vanwoerden-97} (see their Fig.~4). Several
possibilities must be taken into account before one can be convinced that
the lack of an optical HVC absorption feature indicates that the star is in
front of the HVC, and sets a significant lower distance limit. HVCs have
been shown to have large column density variations over small (arcmin)
angular scales \citep[e.g.][]{wakker-schwarz-91-discussion}.  Since HVCs
are usually identified in \HI\ surveys with large beam sizes,
high-resolution (interferometric) radio imaging is necessary, to ensure the
line-of-sight does not pass through a ``sucker hole'' in the cloud, missing
the expected high column density gas.  If the alignment is favorable, we
also require sufficiently high metallicity intervening gas (a typical value
for HVCs is $\sim 0.1 - 0.3$ solar; see
\citet{wakker-01-distance-metallicity}); clearly if this is not the case,
there will be no metal ions to create an absorption line.  Further,
knowledge of the gas metallicity comes from other lines-of-sight, so we
must assume that there are no abundance variations within the cloud.

Optical absorption line studies are often conducted using the \ca~H and K
lines ($\lambda\lambda 3968.469$\,\AA\ and 3933.663\,\AA\ respectively).
These two resonance lines have high oscillator strengths and thus should be
easily detectable.  \citet{prata-wallerstein-67-CaII-NaI} were the first to
specifically target the HVCs (unsuccessfully) using absorption line
studies, with \citet{kepner-68-catalogue} producing the first catalogue of
early-type stars aligned with HVCs. Despite almost 40 years of efforts by
various groups, only a handful of stellar absorption line studies have
successfully detected HVC absorption features\footnote{There is also a
large body of ultraviolet absorption line detections against stellar and
extra-galactic sources.  These ionized high-velocity clouds in some cases
align with \HI\ HVCs, in some cases not.  See e.g.
\citet{sembach-etal-03-OVI-HVC} for details.}, providing firm upper
limits. \citet{vanwoerden-etal-99-nature} detected HVC ``Chain A'' in \ca\
absorption in the optical spectrum of the RR~Lyrae star AD~UMa, giving a
maximum distance of 10\kpc. A few years earlier,
\citet{danly-etal-93-ComplexM} used the {\it IUE} satellite to detect the
HVC Complex~M in the UV resonance lines \ion{O}{1}, \ion{C}{2} and
\ion{Si}{2} toward the early-type star $\mbox{BD}~+38\degr2182$, limiting
the HVC to $z < 4.4\kpc$.

Field Horizontal Branch (FHB) stars are valuable probes for distance
determination work, for two primary reasons: (i) Their intrinsic luminosity
means that sufficiently high signal-to-noise optical spectra can be
obtained for distant halo FHB stars ($d \geq 20$~kpc), meaning they can act
as useful and interesting discriminators of the various origin scenarios,
and (ii) Being hot stars, their spectra are relatively free of intrinsic
absorption features, making the detection and interpretation of the HVC
absorption features relatively straightforward.  Previous authors have also
successfully employed RR~Lyrae stars \citep{vanwoerden-etal-99-nature} or
bright B-stars in the ultraviolet \citep{danly-etal-93-ComplexM} and optical
\citep{smoker-etal-04-HVC}.

Here we present a table of \num~FHB stars which align with \HI\ gas at high
velocities, and the associated \HI\ spectra and moment maps.  These stars
are ideal for high-resolution optical spectroscopy and distance
determinations.  Section~\ref{sec: selection} gives the details about the
selection of HVC distance probes from the HES FHB catalogue of
\citet[][hereafter C05]{christlieb-etal-05-HES-FHB}.  The catalogue of HVC
distance probes is presented in Section~\ref{sec: catalogue}.  A brief
conclusion is given in Section~\ref{sec: conclusion}. The full catalogue
plus \HI\ spectra and summed intensity moment maps towards each star in the
catalogue are presented in the online edition.

\section{Stellar selection and Matching}
\label{sec: selection}
\begin{figure}
  \epsscale{1.0}
  \plotone{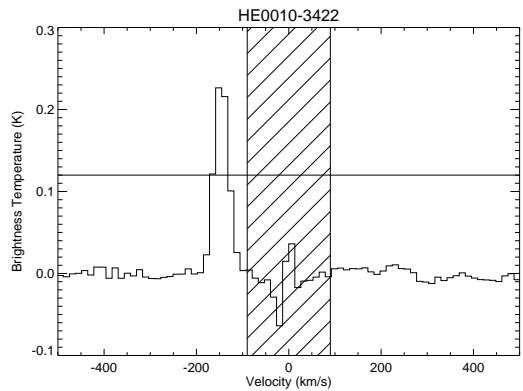}
  \caption{Example \HI\ spectrum of a stellar line-of-sight meeting the
  selection criteria. The horizontal line shows the brightness temperature
  cut-off, while the shaded region indicates velocity selection criteria.
  Note that the HIPASS data is unreliable in the vicinity of the Galactic
  emission near 0\kms. The HIPASS
  spectra are extracted spatially averaging a 3x3 box around the closest
  pixel to the specified coordinate.}
  \label{fig: hipass-spectrum}
\end{figure}

We have taken as our input sample, the C05 catalogue of 8321 FHB stars
selected from the Hamburg/ESO Survey, which has been shown to have
contamination from high-gravity main-sequence stars of $< 16\%$.  For each
stellar line-of-sight, an \HI\ spectrum was extracted from the \HI\
Parkes\footnote{The Parkes telescope is part of the Australia Telescope
which is funded by the Commonwealth of Australia for operation as a
National Facility managed by CSIRO.} All-Sky Survey
\citep[HIPASS;][hereafter P02]{barnes-etal-01-hipass,
putman-etal-02-catalogue} and any emission above a threshold of \jybeam~at
high velocities (\velsel) identified. This threshold corresponds to a
brightness temperature threshold of \tb~for the Parkes Multibeam used for
the HIPASS survey \citep[e.g.][]{putman-etal-03-mshvc} and is set by the
level at which \ca\ absorption lines are expected to be detectable in
optical spectra (see Sec~\ref{sec: optical-detections}).  All
lines-of-sight meeting these criteria were then screened by eye to reject
obvious problems. In all cases, these were due to the wings of emission
features well below the \velsel\ limit extending outside this range above
the level of \tb.  \HI\ moment maps were created from the HIPASS data cubes
around each line of sight, and the positions were matched against several
HVC catalogues.

\subsection{The Hamburg/ESO Survey}

We briefly review the salient points of the HES, while a fuller description
of the survey and its stellar products may be found elsewhere
\citep[e.g.][C05]{wisotzki-etal-96-HES-I}. Originally a survey for bright
quasars, the HES is an objective prism survey which covers a large fraction
of the southern extra-galactic sky ($\delta < +2.5\degr; |b| \gtrsim
30\degr$) to a limiting magnitude of $B \lesssim 18.0$.  As well as finding
quasars, it has also become a valuable resource for identifying many
different classes of interesting stars; e.g. Carbon stars
\citep{christlieb-etal-01-fhlc}, White Dwarfs
\citep{christlieb-etal-01-WD}, metal-poor halo stars
\citep[e.g.][]{Christlieb:2003} and FHB stars (C05). Selection of the FHB
stars was made first on the basis of a color cut.  The second step involved
automated selection criteria based on the summed Balmer line equivalent
widths and the Str{\"o}mgren $c_1$ index. This process resulted in 8321
stars, which have been shown to have a contamination from, e.g. higher
surface gravity A-stars, of $< 16\%$. See C05 for further details.

\subsection{Spectrum Extraction}
\label{sec: spectra}

To identify stars aligned with HVCs, HIPASS \HI\ spectra for all stellar
lines-of-sight were first extracted from the HIPASS HVC cubes. These cubes
were produced from the HIPASS data by P02 using the {\it MINMED5} bandpass
calibration, which recovers much of the extended emission lost in the
original HIPASS bandpass calibration; since the survey was originally
designed for point sources, this loss was not a critical
issue. Nevertheless, when emission extends spatially to fill more than
$\sim 6\degr$ of a full HIPASS $8\degr$ declination scan, the calibration
will not be accurate e.g. for data close to the Galactic plane
\citep[see][for details]{putman-etal-03-mshvc}.

The {\it miriad} task {\it mbspect} was used to extract the spectra,
averaging spatially over a 3x3 pixel box (approximately the HIPASS beam
size of 15.5\,arcmin; HIPASS pixels are 4\,arcmin). {\it mbspect} uses the
nearest pixel to the specified coordinate. All spectra containing emission
above the level of \tb\ at velocities \velsel\ were automatically
flagged. An example of such a spectrum is shown\footnote{Similar figures
are provided for all probes in Table~\ref{tab: catalogue} in Figs.~\fti\ --
\lti\ of the online version.} in Fig~\ref{fig: hipass-spectrum}. We plot
only the velocity range $-500 - +500\kms$ since this is the range occupied
by nearly all Galactic HVCs\footnote{Note that \citet{thilker-etal-04-M31}
have found HVCs associated with M31 at velocities more negative than the
$-500\kms$ limit.}.  The associated moment map is shown in Fig~\ref{fig:
mmap} (see Section~\ref{sec: mmaps} for details).

To create the final FHB/HVC catalogue, the spectra for these \num\
lines-of-sight were then visually inspected to reject obvious noise
artefacts. In some cases, the peak flux was at an absolute velocity less
than the enforced $90\kms$ limit, while significant emission was still
present outside this range. In these cases, stars were retained if the
channel(s) outside the $90\kms$ boundary had a flux comparable to the peak
value. Fig~\ref{fig: hvc-ivc-connection} shows an example of such a
case. In these cases, the peak emission with \velsel\ was recorded.

During screening, it was noted that some spectra contained multiple
high-velocity emission components.  In cases where these were not
immediately identifiable as the same cloud, a separate entry in
Table~\ref{tab: catalogue} was recorded.  These multiple components were
flagged in the catalogue and separate moment maps produced. Obviously,
separate \HI\ spectra are not required, although the spectra are repeated
for completeness.  In all but a few cases, these doubles are flagged as
associated with the Magellanic Stream (see Section~\ref{sec: ms-clouds}).
HE~1008$-$0438 also aligns by chance with the galaxy HIPASS~J1010-04. The
star HE~0053$-$3740 was found to align with a Sculptor Group galaxy and was
excluded. HE~0056$-$4043 shows a noise spike at $\sim -410\kms$ which is
evident in the HIPASS data cubes.

\begin{figure}
  \epsscale{1.0}
  \plotone{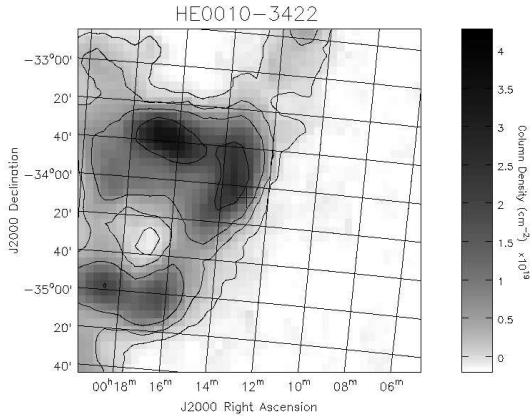}

  \caption{Example moment map of the area around HE~0010$-$3422, the
  line-of-sight shown in Fig~\ref{fig: hipass-spectrum}.  Neutral hydrogen
  contours are drawn at $2.0~\times~10^{18}$, $5.0~\times~10^{18}$,
  $1.0~\times~10^{19}$, $2.5~\times~10^{19}$, $5.0~\times~10^{19}$ and
  $1.0~\times~10^{20} \mbox{cm}^{-2}$ }

  \label{fig: mmap}
\end{figure}

\subsection{Moment Maps}
\label{sec: mmaps}

In order to visualize the spatial locations of the stars with respect to
the gas, summed-intensity moment maps of a region approximately $3\degr
\times 3\degr$ centered around each probe were created. Moment maps were
generated by summing the velocity channel containing the peak flux outside
the \velsel\ boundary and the two adjacent channels.  This summed width
equates to 39.6\kms, which is approximately the 36\kms\ median FWHM of the
HIPASS HVCs (P02) (note, though, that this median is almost certainly an
overestimate of the true value and a product of the coarse velocity
resolution of the HIPASS data). Maps were converted to column densities
using the relation $N(HI) = 1.823\times10^{18} \int T_b dv$
\citep{dickey-lockman-90-MW-HI-review}. Onto these moment maps, contours
were drawn at \HI\ column densities of $2.0~\times~10^{18}$,
$5.0~\times~10^{18}$, $1.0~\times~10^{19}$, $2.5~\times~10^{19}$,
$5.0~\times~10^{19}$ and $1.0~\times~10^{20}$. The positions of the FHB
stars are marked by a ``+'' in the centers of these images, one example of
which is shown\footnote{Similar figures are provided for all probes in
Table~\ref{tab: catalogue} in Figs.~\fti\ -- \lti\ of the online version.}
in Figure~\ref{fig: mmap}.

\subsection{Cloud Matching and the Magellanic Stream}
\label{sec: ms-clouds}
\begin{deluxetable}{cccc}
\tabletypesize{\scriptsize}
\tablecaption{\label{tab: ms-boxes}Magellanic Stream regions.
}
\tablewidth{0pt}
\tablecolumns{6}
\tablehead{
\colhead{Region} &
\colhead{l} &
\colhead{b} &
\colhead{\vlsr}
\\ 
\colhead{} &
\colhead{(deg)} &
\colhead{(deg)} &
\colhead{(\kms)}
}
\startdata
  1 &     --                & $b \leq -70$          & $-300 \leq v \leq 300$  \\
  2 & $270 \leq l \leq 315$ & $-70 \leq b \leq -20$ & $70 \leq v \leq 400$    \\
  3 & $315 \leq l \leq 345$ & $-70 \leq b \leq -50$ & $70 \leq v \leq 200$    \\
  4 & $45  \leq l \leq 110$ & $-70 \leq b \leq -60$ & $-320 \leq v \leq -130$ \\
  5 & $65  \leq l \leq 110$ & $-60 \leq b \leq -48$ & $-390 \leq v \leq -200$ \\
  6 & $65  \leq l \leq 110$ & $-48 \leq b \leq -33$ & $-390 \leq v \leq -250$ \\
\enddata
\tablecomments{Lines-of-sight in these regions, with velocities consistent
with the limits specified, are flagged as associated with the Magellanic
Stream.}
\end{deluxetable}

One of the most prominent features in the southern radio sky at 21\,cm is
the Magellanic Stream (MS). We wanted to differentiate the entries that
probe gas possibly associated with the Stream, from those probing other
HVCs. In this vein we follow the spatial and kinematic regions defined in
Figs.~5 and 7 respectively of \citet{putman-etal-03-mshvc}. Table~\ref{tab:
ms-boxes} shows the regions we adopt as being associated with the Stream.
Probes within these regions are flagged in the catalogue.  Note that exact
definitions of which clouds clouds are associated with the Stream are
difficult and somewhat subjective -- the definitions here should be
considered as a guide only. Further, there is some overlap between the
Stream as defined here and the Sculptor group of galaxies. Despite this
attempted flagging, some of the CHVCs and semi-compact HVCs catalogued by
P02 and \citet[][hereafter D02]{deheij-etal-02-LDS-HVC-cat} are also
flagged as Stream-related.  Since their origin has been proposed to be
quite different from that of the large complexes, and given the imperfect
nature of the MS flagging, these probes may prove quite interesting and are
not excluded from Table~\ref{tab: catalogue}.

Using the HVC catalogues published by PO2, D02 and \citet[][hereafter
WvW91]{wakker-vanwoerden-91-III-catalogue} we have attempted to identify
which clouds are associated with each detection.  From P02, we use both the
tables of HVCs, plus their table of XHVCs -- clouds which have emission at
high velocities, but which clearly link to emission at lower deviation
velocities\footnote{Deviation velocity is the degree to which a velocity
deviates from the maximum allowed by Galactic rotation in that direction
\citep{wakker-91-II-distribution}.}. Since clouds have arbitrary shapes,
this is a difficult process.

In general, the optimal solution to the problem of associating stellar
lines-of-sight with individual clouds would be to obtain component lists
for all clouds in the desired catalogue (i.e. which pixels make up a
particular cloud). Unfortunately, such lists are not available for the
catalogues we searched, so another strategy was required.

From the tables of HVCs and XHVCs reported in P02, we took a position and
angular size.  Assuming a circular cloud, we found all clouds which
encompass the stellar position. We then required that the velocity measured
in the HIPASS spectrum and the velocity of the cloud differ by no more than
the FWHM of the cloud.  For the few cases where multiple clouds met these
criteria, the data cubes were examined and it was immediately apparent
which was the correct entry.

The D02 catalogue is very similar in structure; indeed, the same algorithm
was used by P02 to create the HIPASS catalogue. The one exception is that
D02 do not publish the size of individual clouds, based on the number of
associated pixels, as do P02. They do, however, provide the major and minor
axes of an ellipse which is fit to half the peak flux level.  We used the
major axis length to search a circular area around the reported cloud
position. To avoid associating some clouds with large fractions of the sky
(i.e. some clouds have major axes up to $\sim 165\degr$), we limited the search
radius to 10\,deg. The same FWHM velocity criteria as above were also applied.

For the clouds in the catalogue of WvW91 the same process was repeated.
Since those authors do not tabulate the FWHM for individual clouds, we
assumed a FWHM of 36\kms, based on the global properties of the HIPASS HVCs
listed in Table~2 of P02.

\begin{figure}
  \epsscale{1.0}
  \plotone{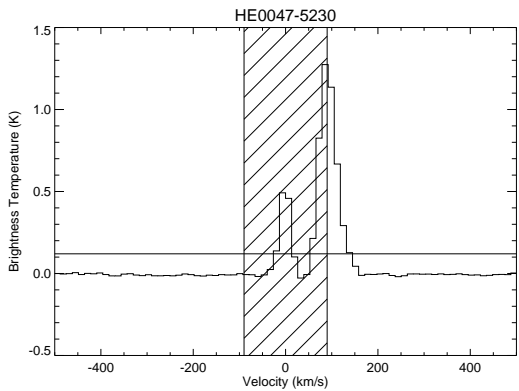}

  \caption{Example \HI\ spectrum of a stellar line-of-sight with peak flux
  inside the \velsel\ boundary. Note the clean separation from the Galactic
  emission at low velocity.}

  \label{fig: hvc-ivc-connection}
\end{figure}

\subsection{Stellar Velocities}

\begin{figure}
  \includegraphics[angle=-90,scale=0.33]{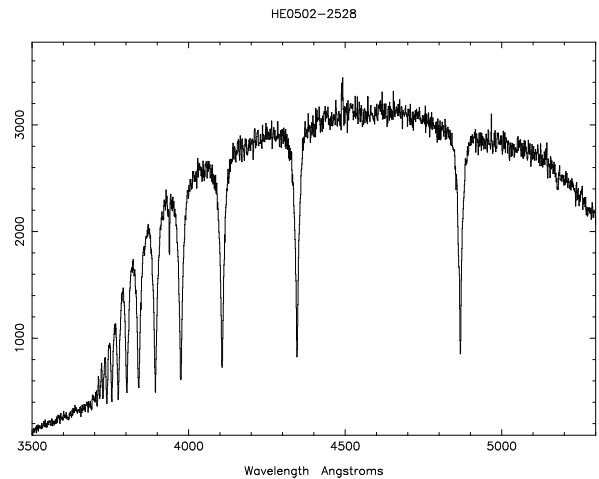}

  \caption{Medium resolution spectrum of the FHB star HE~0502$-$2528. Note
  the general lack of metal absorption lines, which could confuse the
  interpretation of any potential HVC.}

  \label{fig: dbs-spec}
\end{figure}

It will be difficult or impossible to detect HVC absorption features in
high resolution optical spectra if the velocity of the gas and stellar
radial velocity (RV) are too close.  In this case, stellar absorption will
dominate the spectrum and the HVC feature will be lost. As part of a
separate program, we have obtained medium-resolution spectra of 18 stars in
our catalogue using the Double-Beam Spectrograph on the SSO 2.3m. An
example of such a spectrum is shown in Fig.~\ref{fig: dbs-spec}. Note the
general lack of metal absorption lines, with the exception of the stellar
\ca~K line at 3933.663\,\AA.  From these spectra we obtained radial
velocities by fitting Sersic profiles to the \hb, \hc\ and \hd\ Balmer
lines.  Table~\ref{tab: rvs} lists the average stellar RV corrected to the
Local Standard of Rest (LSR) frame, along with the \HI\ velocity and the
absolute difference of the two.  Clearly, the greater the difference
between the stellar and HVC features, the easier a detection will be and we
suggest that a velocity difference of greater than $\sim 50\kms$ will be
needed.  Such observations must be of sufficient resolution (i.e. R = 40000
-- 60000, corresponding to $\sim 7.5 - 5 \kms$ at \ca~K) to resolve the
separation between the HVC, stellar and disk absorption features.

We have performed a simple simulation to estimate, to first-order, the
percentage of halo stars that are expected to have velocities within
50\kms\ of an HVC.  Stars in the halo are drawn from a population with a
total dispersion of $\sigma \simeq 200\kms$.  Assuming an isotropic
distribution of velocities, a single projection (e.g. the line-of-sight
component) will then follow a Gaussian distribution with $\sigma_{LOS}
\simeq 200/\sqrt{3} \simeq 120\kms$. We randomly generated stellar
velocities according to such a distribution.  For each of these velocities,
we also generated a random HVC velocity using the HIPASS HVC velocity
distribution \citep[e.g. Fig~10 of][]{putman-etal-02-catalogue}. Comparing
these two velocities, we find that they are within 50\kms\ of each other
$\sim 13\%$ of the time. It should be noted, however, that this is a global
average; clearly HVCs with lower velocities will have, on average, a chance
velocity alignment more often, while very high velocity HVCs will suffer
such problems less than the average.

\subsection{Optical Detections}
\label{sec: optical-detections}
In order to detect an absorption line in the optical, there must be a
sufficient number of absorbing atoms along the line-of-sight. Assuming an
optically thin gas, \citet{savage-sembach-96-review} give a relation
between the column density $N$ and the equivalent width of the absorption
line $W_\lambda$. Since an \HI\ column density can be measured from the
HIPASS spectra by integrating over the HVC emission line, if the ratio of
$\mbox{Ca}^+$ ions to \ion{H}{1} atoms, \mbox{N(Ca$^+$)/N(\ion{H}{1}}) is
known, we can predict the equivalent width of the \ca\ absorption
features. It should be noted that an accurate determination of
\mbox{N(Ca$^+$)/N(\ion{H}{1}}) is important, since this quantity can vary
by up to 2 orders of magnitude from cloud to cloud
\citep{wakker-01-distance-metallicity}\footnote{However
\citet{wakker-mathis-00-abundance-NHI} show a relation between the
\ion{Ca}{2} abundance and column density with much less variation at a
given column density.}. For the lowest column density features in
Table~\ref{tab: catalogue}, the expected equivalent width is $\sim
20$\,m\AA, which should be detectable using large telescopes in a
reasonable amount of time.

\begin{deluxetable}{lrcr}
\tabletypesize{\scriptsize}
\tablecaption{\label{tab: rvs}Stellar velocity comparisons}
\tablewidth{8cm}
\tablehead{
\colhead{HE name} & 
\colhead{Radial Velocity} &
\colhead{\vlsr} &
\colhead{$v_{diff}$}
\\
\colhead{} &
\colhead{(\kms)} &
\colhead{(\kms)} &
\colhead{(\kms)} 
}
\startdata
HE 0003$-$5248 &   +058   &  +112   &  $-$054  \\
HE 0046$-$3255 &   $-$012 &  +112   &  $-$124  \\
HE 0056$-$3330 &   +113   &  $-$191 &  +304  \\
HE 0157$-$5126 &   +047   &  +165   &  $-$118  \\
HE 0420$-$5405 &   +070   &  +099   &  $-$029  \\
HE 0502$-$2528 &   +370   &  +099   &  +271  \\
HE 0504$-$2509 &   +066   &  +112   &  $-$046  \\
HE 0506$-$2530 &   +204   &  +125   &  +079  \\
HE 0932$-$0404 &   +257   &  +152   &  +105  \\
HE 1004$-$0516 &   +219   &  +112   &  +107  \\
HE 1027$-$2625 &   +261   &  +218   &  +043  \\
HE 1045$-$2550 &   +125   &  +178   &  $-$053  \\
HE 1318$-$2555 &   $-$035 &  +112   &  $-$147  \\
HE 1323$-$2511 &   +182   &  +112   &  +070  \\
HE 2155$-$2243 &   $-$098 &  $-$139 &  +041  \\
HE 2200$-$2222 &   +141   &  $-$099 &  +240  \\
HE 2242$-$6106 &   +094   &  +112   &  $-$018  \\
HE 2304$-$3858 &   $-$038 &  $-$112 &  +074  \\
\enddata 

\tablecomments{Stellar velocity comparisons for stars with medium
resolution optical observations. Listed are the LSR stellar and \HI\
velocities.  A minimum separation of 50\kms\ is suggested for a star to be
useful for high resolution optical observations.  Stellar radial velocities
have been corrected to the LSR frame using the IRAF task {\it rvcorrect},
which assumes a solar motion of 20\kms\ towards $\alpha$ = 18:00:00,
$\delta$ = 30:00:00 (B1900)}

\end{deluxetable}

\section{The Catalogue}
\label{sec: catalogue}
After these selection criteria were applied, we had a total of \num~stars
probing sufficiently dense HVC gas.  These stars are listed in
Table~\ref{tab: catalogue}.  The table contains the following columns:
Column (1) gives the running number. Column (2) indicates the HES name of
the star. Multiple components in a single line-of-sight have separate
entries and are numbered. Columns (3) and (4) list the right ascension and
declination (J2000) of the stellar probe, with Columns (5) and (6) showing
the corresponding Galactic coordinates of the star. The large gap between
RA $\sim$ 13:30 -- 21:30 is due to the HES avoidance of the Galactic
plane. Column (7) gives the HES B magnitude, which is accurate to
$\sim0.2\,\mbox{mag}$ \citep{wisotzki-etal-00}. The distance to the star
(in kpc) is shown in column (8) and is taken directly from C05. Briefly,
the distance was derived by those authors by estimating the absolute
magnitude $M_{\rm v}$ from the \bv\ color.  A metallicity correction was
applied based on an assumed average halo metallicity and the distance
moduli computed. For any higher gravity A-star contaminants, this distance
will be incorrect. The peak brightness temperature (in K) is shown in
column (9); P02 quote a 5$\sigma$ sensitivity of 0.04\,K. Column (10) shows
the \HI\ column density along the line-of-sight, which was measured from
the moment maps. The LSR velocity (\kms) of the peak \HI\ emission outside
the \velsel\ zone is given in column (11). This is the velocity of the
HIPASS channel containing the peak flux -- no attempt at profile fitting
was made. The HIPASS velocity resolution is 26.4\kms\ after hanning
smoothing.  An asterisk in column (12) indicates an association with the
Magellanic Stream, (see Sec.~\ref{sec: ms-clouds}).  The matches to the P02
HVCs and XHVCs, the WvW91 HVCs and the D02 HVCs are given in columns (13),
(14), (15) and (16) respectively. For the P02 and D02 catalogues, the
original nomenclature classifying the cloud (e.g. HVC, CHVC, HVC: etc) has
been retained.

\section{Conclusion}
\label{sec: conclusion}
We have presented a catalogue of FHB stars, selected from the Hamburg/ESO
Survey, that fortuitously align with high-velocity \HI\ gas.  These stars
should make ideal candidates for echelle spectroscopy.  Such observations
offer the possibility of detecting the doppler-shifted absorption features
in the spectrum, placing an upper limit on the distance to the HVC.  In a
similar fashion, non-detections place lower limits on the distance.
Distance limits will help to resolve the question of the origins of HVCs
and their role in the formation and evolution of Galaxies.

\acknowledgments 

The financial support of the Australian Research Council, through its
Discovery Project and Linkage International Award schemes, is gratefully
acknowledged. We thank the scientific editor, W. Butler Burton, for his
extensive and helpful suggestions.

\bibliography{BHB,nc,HVC,HVC-FUSE,HES,me,wakker,mary,HI,HVC-galfountain,CHVC,ISM,chemical-evolution,hipass,Halo}
\bibliographystyle{apj}

\clearpage

\begin{deluxetable}{llcccccccccccccl}
\tabletypesize{\tiny}
\tablecaption{\label{tab: catalogue}Catalogue of FHB stars aligned with HVCs.}
\tablewidth{0pt}
\tablehead{
\colhead{Num} & 
\colhead{HEname} & 
\colhead{RA (J2000)} & 
\colhead{DEC (J2000)} & 
\colhead{$l$} &
\colhead{$b$} & 
\colhead{B mag} & 
\colhead{Distance} & 
\colhead{$T_b$} &
\colhead{N(\ion{H}{1})} & 
\colhead{\vlsr} &
\colhead{MS Flag} & 
\colhead{HVC} &
\colhead{XHVC} &
\colhead{WvW91} & 
\colhead{de Heij} 
\\
\colhead{} & 
\colhead{} & 
\colhead{} & 
\colhead{} & 
\colhead{(deg)} &
\colhead{(deg)} & 
\colhead{(mag)} & 
\colhead{(kpc)} &
\colhead{(K)} & 
\colhead{(cm$^{-2}$)} & 
\colhead{\kms} & 
\colhead{} & 
\colhead{} & 
\colhead{} &
\colhead{} & 
\colhead{} 
}
\startdata
327 & HE2129$-$0202 &   21:32:31.5 & $-$01:49:20 &  52.18 & $-$36.127 & 16.8  & 14.3 & 0.13 & 7.4e+18 & $+$112 & - & CHVC\#425  & -         & -        & HVC\#162   \\
328 & HE2141$-$2039 &   21:44:28.7 & $-$20:25:29 &  31.44 & $-$46.950 & 16.3  &  6.8 & 0.14 & 8.3e+18 & $-$178 & - & HVC\#291   & -         & -        & CHVC\#68   \\
329 & HE2153$-$2323 &   21:56:32.7 & $-$23:09:24 &  28.70 & $-$50.416 & 14.5  &  4.8 & 0.25 & 1.6e+19 & $-$112 & - & HVC\#283   & -         & -        & HVC\#51    \\
330 & HE2155$-$2133 &   21:58:25.9 & $-$21:18:47 &  31.67 & $-$50.320 & 16.7  & 11.9 & 0.20 & 1.2e+19 & $-$125 & - & HVC\#283   & -         & -        & HVC\#45    \\
331 & HE2155$-$2243 &   21:58:39.9 & $-$22:29:29 &  29.92 & $-$50.711 & 15.3  &  6.2 & 0.77 & 4.8e+19 & $-$139 & - & HVC\#283   & -         & -        & CHVC:\#64  \\
332 & HE2200$-$2222 &   22:02:54.8 & $-$22:08:06 &  30.90 & $-$51.551 & 16.4  & 10.4 & 0.39 & 2.3e+19 & $-$099 & - & HVC\#283   & -         & -        & HVC\#45    \\
333 & HE2200$-$2157 &   22:03:17.4 & $-$21:42:31 &  31.59 & $-$51.528 & 17.7  & 21.5 & 0.25 & 1.5e+19 & $-$099 & - & HVC\#283   & -         & -        & HVC\#45    \\
334 & HE2200$-$2249 &   22:03:32.4 & $-$22:35:13 &  30.26 & $-$51.824 & 15.7  &  7.2 & 0.13 & 7.7e+18 & $-$099 & - & HVC\#283   & -         & -        & HVC\#45    \\
335 & HE2235$-$4908 &   22:38:52.5 & $-$48:52:38 & 343.54 & $-$56.474 & 16.2  &  9.3 & 0.22 & 1.4e+19 & $+$099 & * & :HVC\#1877 & -         & -        & -          \\
336 & HE2236$-$6040 &   22:39:26.1 & $-$60:25:07 & 327.61 & $-$49.916 & 16.7  & 14.8 & 0.17 & 1.1e+19 & $+$099 & - & HVC\#1727  & -         & -        & -          \\
337 & HE2242$-$6106 &   22:45:48.5 & $-$60:51:02 & 326.29 & $-$50.207 & 14.0  &  3.9 & 0.22 & 1.1e+19 & $+$112 & * & HVC\#1727  & -         & -        & -          \\
338 & HE2247$-$5811 &   22:50:18.7 & $-$57:55:22 & 329.10 & $-$52.619 & 16.3  & 12.5 & 0.15 & 8.4e+18 & $+$099 & * & -          & XHVC\#147 & -        & -          \\
339 & HE2248$-$5733 &   22:51:17.5 & $-$57:17:54 & 329.72 & $-$53.132 & 14.5  &  5.4 & 0.21 & 1.2e+19 & $+$099 & * & -          & XHVC\#147 & -        & -          \\
340 & HE2249$-$5740 &   22:52:46.8 & $-$57:24:30 & 329.36 & $-$53.204 & 15.1  &  6.4 & 0.49 & 2.8e+19 & $+$099 & * & -          & XHVC\#147 & -        & -          \\
341 & HE2251$-$5755 &   22:54:18.6 & $-$57:39:02 & 328.83 & $-$53.197 & 17.1  & 16.7 & 0.50 & 2.7e+19 & $+$099 & * & -          & XHVC\#147 & -        & -          \\
342 & HE2252$-$5943 &   22:55:25.1 & $-$59:27:39 & 326.53 & $-$52.025 & 16.4  & 12.6 & 0.20 & 1.1e+19 & $+$112 & * & -          & XHVC\#147 & -        & -          \\
343 & HE2253$-$6056 &   22:56:35.4 & $-$60:40:07 & 325.02 & $-$51.258 & 16.5  & 12.6 & 0.20 & 1.3e+19 & $+$125 & * & HVC\#1727  & XHVC\#147 & -        & -          \\
344 & HE2254$-$5929 &   22:57:17.7 & $-$59:13:38 & 326.52 & $-$52.366 & 16.2  & 10.7 & 0.36 & 2.1e+19 & $+$112 & * & -          & XHVC\#147 & -        & -          \\
345 & HE2255$-$5806 &   22:58:39.1 & $-$57:50:29 & 327.92 & $-$53.470 & 16.7  & 13.0 & 0.17 & 9.7e+18 & $+$152 & * & -          & XHVC\#147 & -        & -          \\
346 & HE2256$-$5859 &   22:59:12.0 & $-$58:43:37 & 326.80 & $-$52.890 & 16.8  & 13.4 & 0.30 & 1.7e+19 & $+$138 & * & -          & XHVC\#147 & -        & -          \\
347 & HE2301$-$6154 &   23:04:49.1 & $-$61:38:18 & 322.80 & $-$51.186 & 15.9  &  9.4 & 0.34 & 1.8e+19 & $+$099 & * & -          & -         & -        & -          \\
348 & HE2302$-$6140 &   23:05:35.6 & $-$61:24:04 & 322.93 & $-$51.422 & 15.5  &  7.6 & 0.60 & 3.4e+19 & $+$099 & * & -          & -         & -        & -          \\
349 & HE2304$-$3858 &   23:07:26.3 & $-$38:41:52 & 358.27 & $-$65.426 & 16.4  &  9.8 & 0.42 & 2.5e+19 & $-$112 & - & CHVC\#1976 & -         & WvW\#548 & -          \\
350 & HE2305$-$6146 &   23:08:24.6 & $-$61:30:00 & 322.41 & $-$51.551 & 16.9  & 16.3 & 0.56 & 3.3e+19 & $+$099 & * & -          & -         & -        & -          \\
351 & HE2307$+$0131 &   23:09:46.4 & $+$01:47:55 &  78.63 & $-$52.111 & 17.4  & 18.1 & 0.19 & 1.1e+19 & $-$310 & - & HVC\#476   & -         & -        & HVC\#265   \\
352 & HE2307$+$0125 &   23:09:58.4 & $+$01:41:38 &  78.59 & $-$52.227 & 16.9  & 14.1 & 0.15 & 9.1e+18 & $-$310 & - & HVC\#476   & -         & -        & HVC\#265   \\
353 & HE2308$-$3256 &   23:11:37.6 & $-$32:39:58 &  12.50 & $-$67.736 & 17.5  & 20.3 & 0.16 & 8.9e+18 & $-$165 & - & CHVC\#129  & -         & -        & -          \\
\enddata

\tablecomments{Table~\ref{tab: catalogue} is published in its entirety in
the electronic edition of the {\it Astrophysical Journal}. A portion is
shown here for guidance regarding its form and content.}
\end{deluxetable}
\clearpage

\end{document}